\documentclass[preprint,12pt,margin=1in]{elsarticle}
\usepackage[margin= 1in]{geometry}
\usepackage{amsmath}
\usepackage{tabularx}
\usepackage{graphicx}
\usepackage{adjustbox}
\usepackage{multirow}
\usepackage{subcaption}
\usepackage{multirow}
\usepackage{booktabs}
\usepackage[numbers]{natbib}

\usepackage{amssymb}

\journal{Journal of Magnesium and Alloys}

\begin{document}

\begin{frontmatter}



\title{Mechanical Property Design of Bio-compatible Mg alloys 
using Machine-Learning Algorithms}


\author[inst1]{Parham Valipoorsalimi}

\affiliation[inst1]{McGill University={Department of Mining and Materials Engineering},
            addressline={McGill University}, 
            city={Montreal},
            state={Quebec},
            country={Canada}}
\author[inst2]{Yuksel Asli Sari}

\affiliation[inst2]{Queen's University={The Robert M. Buchan Department of Mining},
            addressline={Queen's University}, 
            city={Kingston},
            state={Ontario},
            country={Canada}}
\author[inst1]{Mihriban Pekguleryuz}


\begin{abstract}
Magnesium alloys are attractive options for temporary bio-implants because of their biocompatibility, controlled corrosion rate, and similarity to natural bone in terms of stiffness and density. Nevertheless, their low mechanical strength hinders their use as cardiovascular stents and bone substitutes. While it is possible to engineer alloys with the desired mechanical strength, optimizing the mechanical properties of biocompatible magnesium alloys using conventional experimental methods is time-consuming and expensive. Therefore, Artificial Intelligence (AI) can be leveraged to streamline the alloy design process and reduce the required time. In this study, a machine learning model was developed to predict the yield strength (YS) of biocompatible magnesium alloys with an $R^2$ accuracy of 91\%. The predictive model was then validated using the CALPHAD technique and thermodynamics calculations. Next, the predictive model was employed as the fitness function of a genetic algorithm to optimize the alloy composition for high-strength biocompatible magnesium implants. As a result, two alloys were proposed and synthesized, exhibiting YS values of 108 and 113 MPa, respectively. These values were substantially higher than those of conventional magnesium biocompatible alloys and closer to the YS and compressive strength of natural bone. Finally, the synthesized alloys were subjected to microstructure analysis and mechanical property testing to validate and evaluate the performance of the proposed AI-based alloy design approach for creating alloys with specific properties suitable for diverse applications.
\end{abstract}




\end{frontmatter}

\section{Introduction}

Biomaterials are specifically designed to replace all or part of an absent organ within the body \citep{migonney2014history}. There are four main types of biomaterials: ceramics, metals, polymers, and composites. Compared to ceramics or polymeric materials, metals are often better suited for applications such as artificial joints, bone plates and screws, and dental root implants due to their high mechanical strength and fracture toughness \citep{wang2004bioactive}. The most commonly used metallic biomaterials are stainless steels, titanium (Ti) alloys, and cobalt-chrome-based alloys \citep{zeng2008progress}. However, releasing toxic metallic ions or particles due to corrosion or wear processes is a drawback of these metallic biomaterials. Additionally, because the elastic moduli of current metallic biomaterials do not match those of natural bone tissue, they can have stress-shielding effects that reduce the stimulation of bone remodeling and new bone growth, lowering implant stability \citep{staiger2006magnesium}. Biodegradable implants have emerged as an alternative to conventional implants because they provide temporary support until the bone heals and then degrade gradually after bone healing.

Currently, the design and use of biodegradable metals are in the spotlight after decades of improving corrosion-resistant metallic biomaterials. In the ideal case, the biodegradable implant materials disappear as the tissue fully recovers, eliminating the need for repeat interventions for implant removal. When a biodegradable material successfully promotes tissue healing without leaving any implant residues, it should gradually degrade in vivo, with an appropriate host response to released corrosion products. Materials should be non-toxic or composed of metallic elements that the human body can metabolize. As a result, biodegradable alloys based on magnesium are promising materials for medical applications \citep{lu2021biodegradable}.

The use of magnesium as a biomaterial dates back to 1878 when the physician Edward C. Huse utilized pure magnesium wires as ligatures to stop bleeding vessels in three human patients. He had already discovered that magnesium corrosion was slower in vivo, and the time required for complete degradation was proportional to the size of the magnesium wire used \citep{witte2010history}.

Recent research has focused on improving the bio-corrosion behaviour and mechanical integrity of Mg implant materials, although biocompatibility has not always been prioritized \citep{erbel2007temporary}. In a study on Mg-based bone implants, researchers examined the corrosion rates, ductility, and fracture behaviour of various alloys, including Mg-0.8Ca, WE43 (Mg-4Y-3RE, where RE = rare earth elements), and LAE442 (Mg-4Li-4-2RE) \citep{erbel2007temporary}. The WE43 alloy had the lowest corrosion rates but unpredictable fracture behaviour, while Mg-0.8Ca showed promising biocompatibility, similar corrosion rates to WE43, and good ductility but low strength. The LAE442 alloy had acceptable mechanical properties, but its rare earth element accumulation in the embedding tissue raised concerns about its biocompatibility, and the surrounding tissue showed some accumulation of Al. Subsequent investigations confirmed that specific levels of Ca and Ca+Sr improve the bio-degradation resistance of Mg while maintaining bio-compatibility \citep{wan2008preparation,you2000effect}. Zn was also found to enhance bio-corrosion resistance \citep{berglund2012synthesis}. Researchers have developed Mg alloys for cardiovascular stents that have improved biocompatibility and form a Sr-modified hydroxyapatite layer when in contact with body fluids or simulated body fluid, benefiting bone growth and healing \citep{bornapour2014magnesium,bornapour2013biocompatibility,bornapour2015thermal,top2017characteristics}. However, the mechanical performance of these alloys needs optimization before they can be used more broadly \citep{chen2018mechanical}.

This study focuses on optimizing the mechanical performance of biocompatible Mg alloys for bone implants. Tab.~\ref{tab:properties} compares the properties of commonly used biomaterials and natural bone. It shows that the elastic modulus of Mg is much closer to that of natural bone compared to other biomaterials such as Ti, stainless steel, and Co-Cr alloys. The similar elastic modulus of Mg to bone can prevent stress transfer from bone to the implant, also known as stress shielding, which occurs when the mechanical properties of bone and implant are different, specifically when the bone properties are lower \citep{niinomi2011titanium}. Although the bone strength is lower than that of bio-inert metals currently used, it is higher than that of biodegradable polymers.



One of the ways mechanical properties can be improved in metallic materials is via alloying to create a solid solution and/or second-phase strengthening \citep{callister2018materials}. However, these strengthening mechanisms are challenged in developing bio-compatible metallic alloys due to the limitation of using only bio-compatible element additions. Heat treatment is another way to enhance the mechanical characteristics of Mg alloys assuming that the solubility of a specific alloying element increases with temperature \citep{chen2018mechanical}. Trial/error is the current primary method of finding promising alloys for different applications. Due to the vast search space and the unknown, complex relationships between attributes, this method is costly, and it can take many years to produce an alloy for a specific application. The present study offers machine learning as a promising method for alloy design. 

\begin{table}[htb]
\caption{Physical and mechanical properties of various implant materials to natural bone \citep{staiger2006magnesium,kundu2014rapid}}
\label{tab:properties}
\centering
\begin{adjustbox}{width=1\textwidth,center=\textwidth}
\begin{tabular}{lcccccc} 
\hline
Properties                       & Natural bone                                                                         & Magnesium alloys & Ti alloys & Co-Cr alloys & Stainless steel & Synthetic hydroxyapatite  \\ 
\hline
Density (g/cm3)                  & 1.8-2.1                                                                              & 1.74-2           & 4.4-4.5   & 8.3-9.2      & 7.9-8.1         & 3.1                       \\
Elastic modulus (Gpa)            & 3-20                                                                                 & 41-45            & 110-117   & 230          & 230             & 73-117                    \\
Compressive yield strength (MPa) & \begin{tabular}[c]{@{}c@{}}106-131 (Transverse)\\131-224 (Longitudinal)\end{tabular} & 23-300           & 758-1117  & 450-1000     & 4500-1000       & 600                       \\
Tensile yield strength (MPa)     & \begin{tabular}[c]{@{}c@{}}51-66~(Transverse)\\78-151~(Longitudinal)\end{tabular}    & 23-300           & 758-1117  & 450-1000     & 4500-1000       & -                         \\
Fracture toughness ($MPaM^{1/2}$) & 3-6                                                                                  & 15-40            & 55-115    & N/A          & N/A             & 0.7                       \\
\hline
\end{tabular}
\end{adjustbox}
\end{table}

Machine learning has already been used for highly complex problems in different fields and has achieved remarkable achievements in computer vision, natural language processing, speech recognition, and financial analysis \citep{nassif2021deep, ebrahimi2016developing, ozbayoglu2020deep}. It has begun to be employed in materials science for high entropy alloy design \citep{zhang2020phase,yang2022machine,ozdemir2022machine}, property prediction \citep{xu2020predicting,zhang2021machine}, physical properties prediction \citep{zhang2018strategy,ramprasad2017machine}, and fatigue life and crack location prediction \citep{rovinelli2018using}.

Xu et al. \citep{xu2020predicting} used an artificial neural network and SVM to predict UTS, YS, and elongation of rolled and extruded AZ31 alloys. However, they observed 5.4\%, 23\%, and 272\% errors when they fabricated the suggested alloy in the lab and compared the properties with the models' predictions. Liu et al. \citep{liu2021accelerated} implemented four different methods on a dataset of alloy compositions and heat treatment conditions to predict the hardness of Mg alloys. XGBoost showed the best accuracy on the mentioned dataset, and active learning \citep{lookman2019active} was used to find the candidates with the highest hardness. Xia et al. \citep{xia2016artificial} researched the prediction of corrosion and hardness of Mg alloys using an artificial neural network. However, the ANN model was built using small amounts of Zn, Ca, Zr, Gd, and Sr with only 53 data points, which decreased the generality of the model. Birbilis et al. \citep{birbilis2011combined} used an artificial neural network to investigate the corrosion rate and YS of Mg-RE alloys with various alloying combinations. Machine learning was also investigated as a method of investigating/predicting the relationship between composition and mechanical characteristics of Mg-Al-Zn alloys by Liu et al. \citep{liu2005model,krupinski2012prediction}.

This study proposes a machine learning framework to predict the mechanical properties of biocompatible Mg alloys and propose elemental compositions with strong mechanical properties. After the development of a number of machine learning models, the accuracy of the best model is further validated using the CALPHAD technique, employing alloy attributes simulated via thermodynamic computations. Subsequently, the proposed machine learning pipeline uses genetic algorithms to optimize the composition of two new alloys as biocompatible and biodegradable implant material with adequate tensile properties.

The use of machine learning techniques in this study provides an effective and comprehensive approach to investigating the complex relationships between composition, second phases, heat treatment conditions, and mechanical properties of biocompatible Mg alloys. Furthermore, the ability to utilize the results of this study to design a new alloy with improved properties highlights the potential of artificial intelligence in facilitating the development of novel materials for diverse applications.

This research makes two original contributions. First, it introduces a new technique to validate AI models using simulation and thermodynamic calculations. Second, the study demonstrates the use of predictive models in conjunction with genetic algorithms to design a new alloy with specific additions and mechanical properties for a particular application, which has not been explored before.

\section{Materials and methods}

\begin{figure*}[htbp]
    \centering
    \includegraphics[width=1\linewidth]{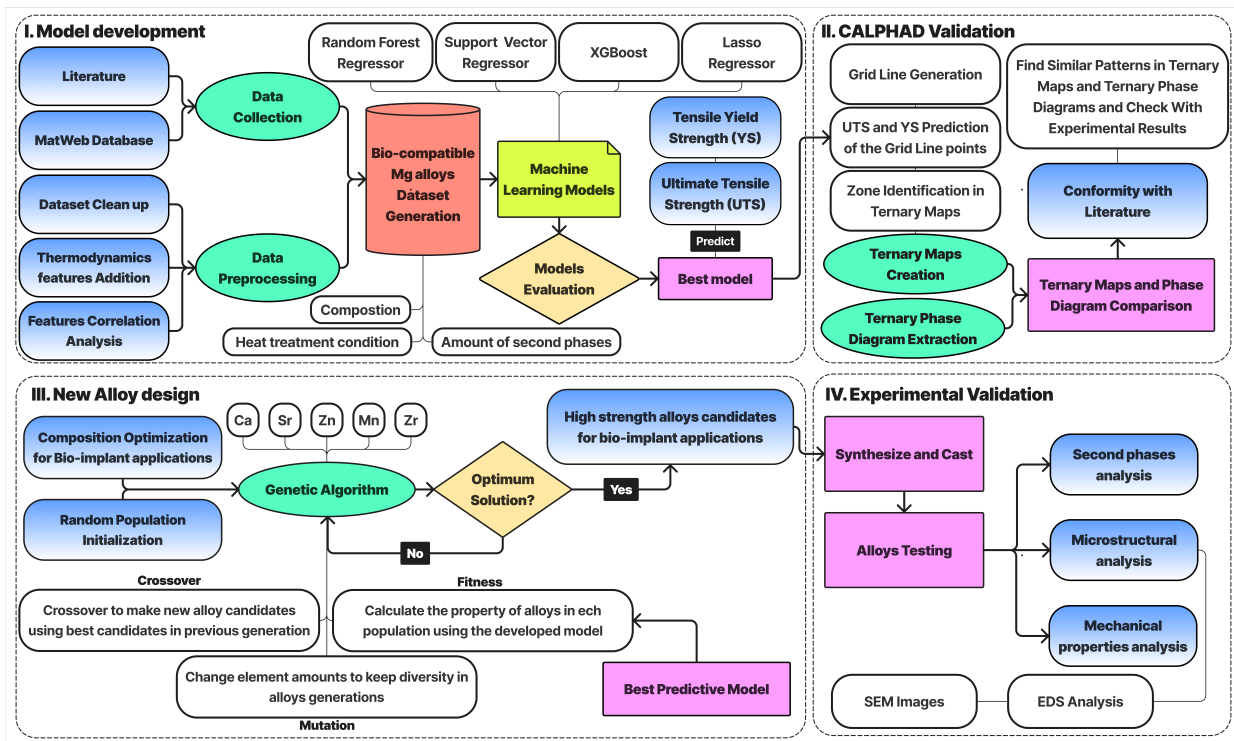} 
\caption{Flowchart of the design and validation process}
\label{fig:flow}
\end{figure*}

The proposed approach comprises four main steps, as depicted in Fig. \ref{fig:flow}. These steps involve developing a predictive model, verifying the model's accuracy using the CALPHAD technique, utilizing genetic algorithms to identify potential alloys, and conducting experimental validation.

\subsection{Software and model Implementation}
\subsubsection{Software}
The predictive models for this project were implemented using the Python programming language and various libraries that facilitate different aspects of the modeling process. Specifically, the Pandas library was used for data analysis, the NumPy library for numerical calculations for the microstructural model and the genetic algorithm, and the Scikit-learn library for implementing machine learning models. Additionally, the FactSage software was utilized to determine the potential amount of second phases at room temperature and to generate ternary phase diagrams for validation by simulation. A more detailed discussion on the construction and implementation of the models can be found in research by \citet{valipoorsalimi_2023}.

\subsubsection{Data collection and preprocessing}
\begin{figure}[htbp]
    \centering
    \includegraphics[width=1\linewidth]{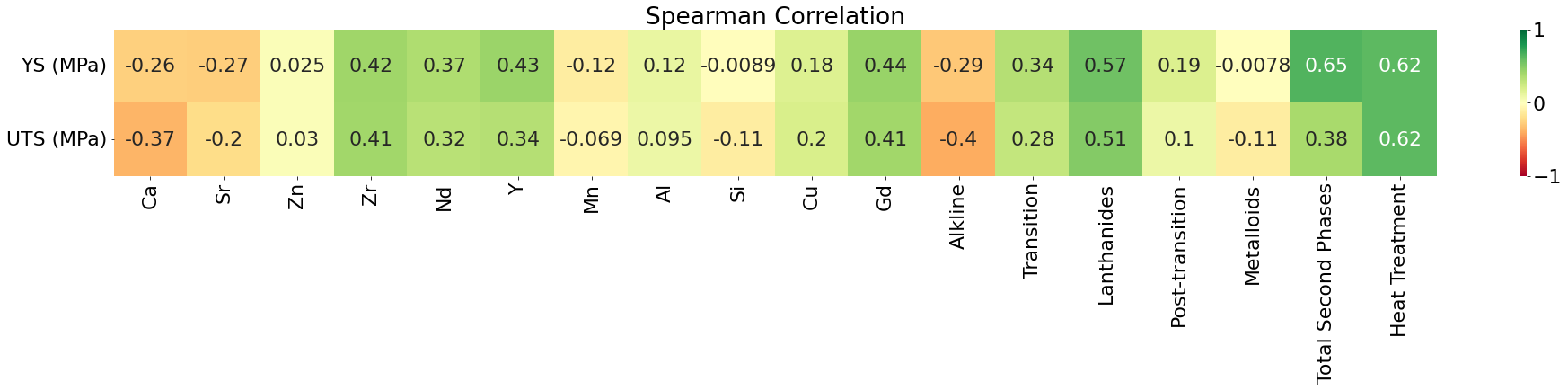} 
\caption{Spearman correlation analysis of the final dataset}
\label{fig:corr}
\end{figure}

A primary dataset was compiled from various literature sources and the MatWeb online database to train the machine learning models. Initially, the dataset consisted of 384 data points, including input features such as alloy composition and heat treatment conditions and outputs such as YS and ultimate tensile strength (UTS). After thorough preprocessing and data cleaning, 284 data points were selected to train the machine learning models. The selected data included the alloy composition, heat treatment conditions, and the amount of second phases present in the alloys. Details on the data collection and preprocessing can be found in the study by \citet{valipoorsalimi_2023}. Fig. \ref{fig:corr} displays the correlation between the input features and outputs of the final dataset. It can be observed that heat treatment conditions and the total amount of second phases are the most correlated features with both mechanical properties.

\subsubsection{Random forest model (RF)}
The RF model is a supervised learning algorithm that leverages an ensemble of decision trees to make predictions. Unlike relying on a single predictor, the model creates a "forest" of decision trees, where each tree acts as an individual learner within the ensemble. Each decision tree comprises decision nodes and branches that lead to a prediction based on the alloy's composition, heat treatment conditions, and amount of second phases. To diversify the individual learners, each tree is built using a subset of the training data instead of the entire dataset. Although this may sound counterintuitive and decrease the performance of each decision tree, the relatively uncorrelated errors produced by the trees enhance the ensemble performance. The model's prediction is determined by averaging the predictions of all the trees in the forest.

One significant advantage of the RF model is its ability to handle high-dimensional data accurately, including missing values, categorical variables, and a large number of classes. It also provides an estimate of variable importance, which is helpful in selecting essential features for the model. However, training the model can be time-consuming and computationally expensive, particularly with large datasets. The models produced can also be complex and challenging to interpret. Moreover, if not set up correctly, the model may not generalize well to new data and could overfit in the case of noisy data.

\subsubsection{Genetic Algorithm (GA)}
GA is a population-based algorithm in which each parameter represents a gene, and each solution corresponds to a chromosome. GA uses a fitness (objective) function to assess each chromosome's fitness within the population \citep{mirjalili2019evolutionary}.

This study uses GA to investigate new compositions with different heat treatment conditions in the search space. The fitness function corresponds to YS and UTS, so the most promising alloys can be explored according to the previously built RF model. For this algorithm, each alloy is considered a chromosome, each alloying element is referred to as a gene, and the set of alloys in each iteration is the population. Fig.~\ref{fig:GA-method} indicates the mentioned assumptions. Elements 1 through 11 (El 1 -- El 11) are the alloying elements that can be altered within a specified range, and HT is the heat treatment condition, which is 0 for alloys that have not been heat treated and 1 for alloys that have been heat treated.

The four stages listed below were mainly taken into account in this research when implementing GA:

\begin{enumerate}
    \item A population of 20 possible random alloys is initialized to uniformly distribute the solutions across the search space to increase population diversity and improve the likelihood of identifying viable locations.
    
    \item The best model from the previous stage was selected to predict the mechanical properties of alloys in the population and was considered the fitness function. The best individuals are then selected based on their fitness to pass their genes (alloying elements) for the production of the next generation. The probability of occurrence of an individual in the next generation is proportional to their fitness which means that the likelihood of a poor individual contributing to the next generation is very low.
    
    \item Next step is the crossover, in which the chromosomes (alloys composition and heat treatment condition) of two selected parents are combined to make a new chromosome. There are various techniques for crossover in the literature. In this study, a single-point crossover is used. The chromosomes of two parent solutions are switched before and after a single point.
    
    \item Lastly, one or more genes of newly made chromosomes are changed by (0.05). This step increases the randomness, which helps maintain the population's diversity.

\begin{figure}[htb]
    \centering
    \includegraphics[width=0.6\linewidth]{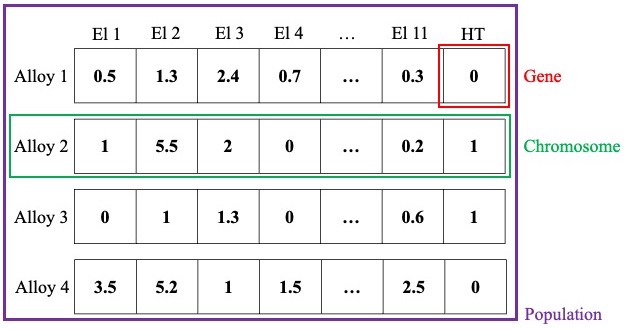} 
\caption{Illustration of population, chromosome, and gene in proposed GA. El 1--El 11 are alloying elements, and HT is heat treatment condition for each alloy in populations}
\label{fig:GA-method}
\end{figure}

\end{enumerate}

\subsection{Alloy synthesis}
\begin{figure}[htb]
    \centering
    \includegraphics[width=0.8\linewidth]{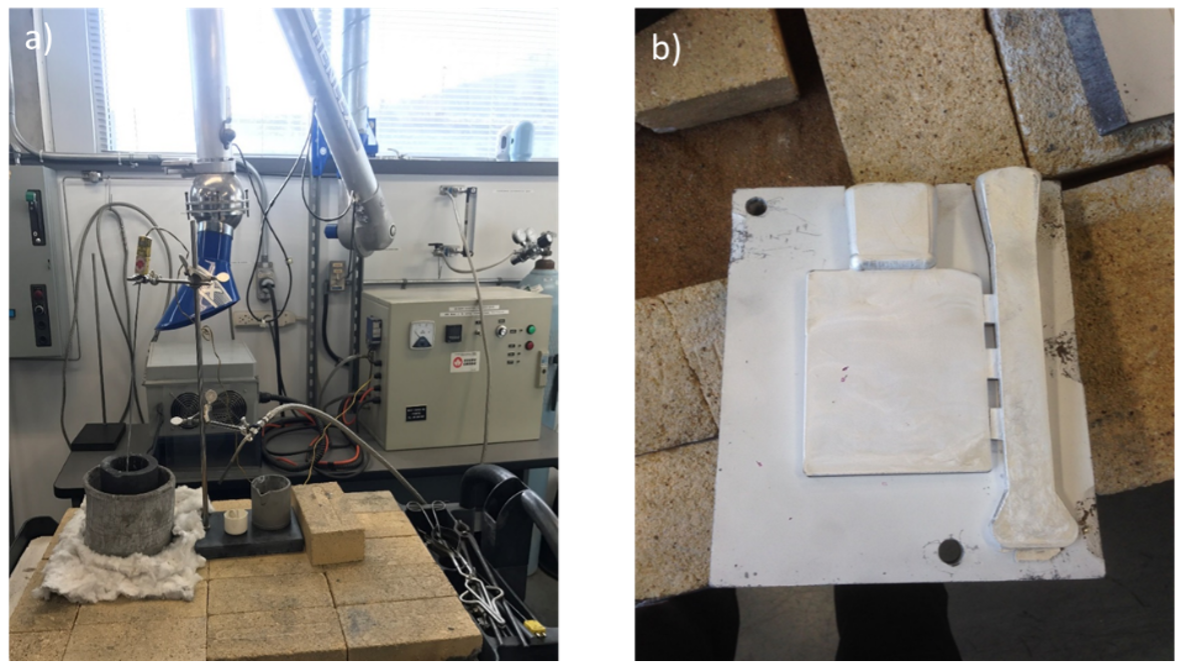} 
\caption{a)  Graphite crucible and the induction furnace used for alloy synthesis, b) permanent-mould cast plate.}
\label{fig:Cast}
\end{figure}

Two Mg alloys suggested by the predictive model were synthesized in the McGill Light Metals Research lab. Pure Mg was melted, and alloying elements were introduced at 700 $^{\circ}$C in a graphite crucible and Norax Canada induction furnace under $CO_{2}$/0.5\% SF6 protective atmosphere. Pure Sr and pure Zn and master alloys of Mg-30 Ca and Mg-0.75 Zr were used, and a recovery rate of 85\% for all element additions was employed. The molten Mg alloy was poured into a steel die coated with a boron-nitride release coating and cooled to produce a plate sample (Fig. \ref{fig:Cast}). Lastly, the chemical compositions were determined using inductively coupled plasma (ICP) analysis.

\subsection{Microstructural analysis}
A Hitachi SU - 3500 Pressure Scanning Electron Microscope (VP-SEM) with energy dispersive spectroscopy (EDS) capability was used to examine the second phases present in the as-cast alloys. The samples were prepared by grinding with 1200 grit SiC papers, polishing with 1-micron diamond paste on low-napped polishing cloths, and finally polishing with colloidal silica to achieve a smooth and mirror-like surface finish. The samples were then analyzed with EDS point analysis and elemental mapping in the BSE (back-scattered electron) mode at an accelerating voltage of 15 kV to assess the chemical composition of the phases and their distribution in the microstructure.

\subsection{Hardness test}
The STRUERS DuraScan-80 machine was used to measure the hardness of 10 points on a material with a load of 200 g and a 15-second dwell period. The points were separated by at least three times the diagonal measure of the indenter; the average value was calculated for an accurate measure of hardness.

\section{Results and Discussion}

\subsection{Predictive Model}
\begin{figure}[htb]
    \centering
    \includegraphics[width=0.7\linewidth]{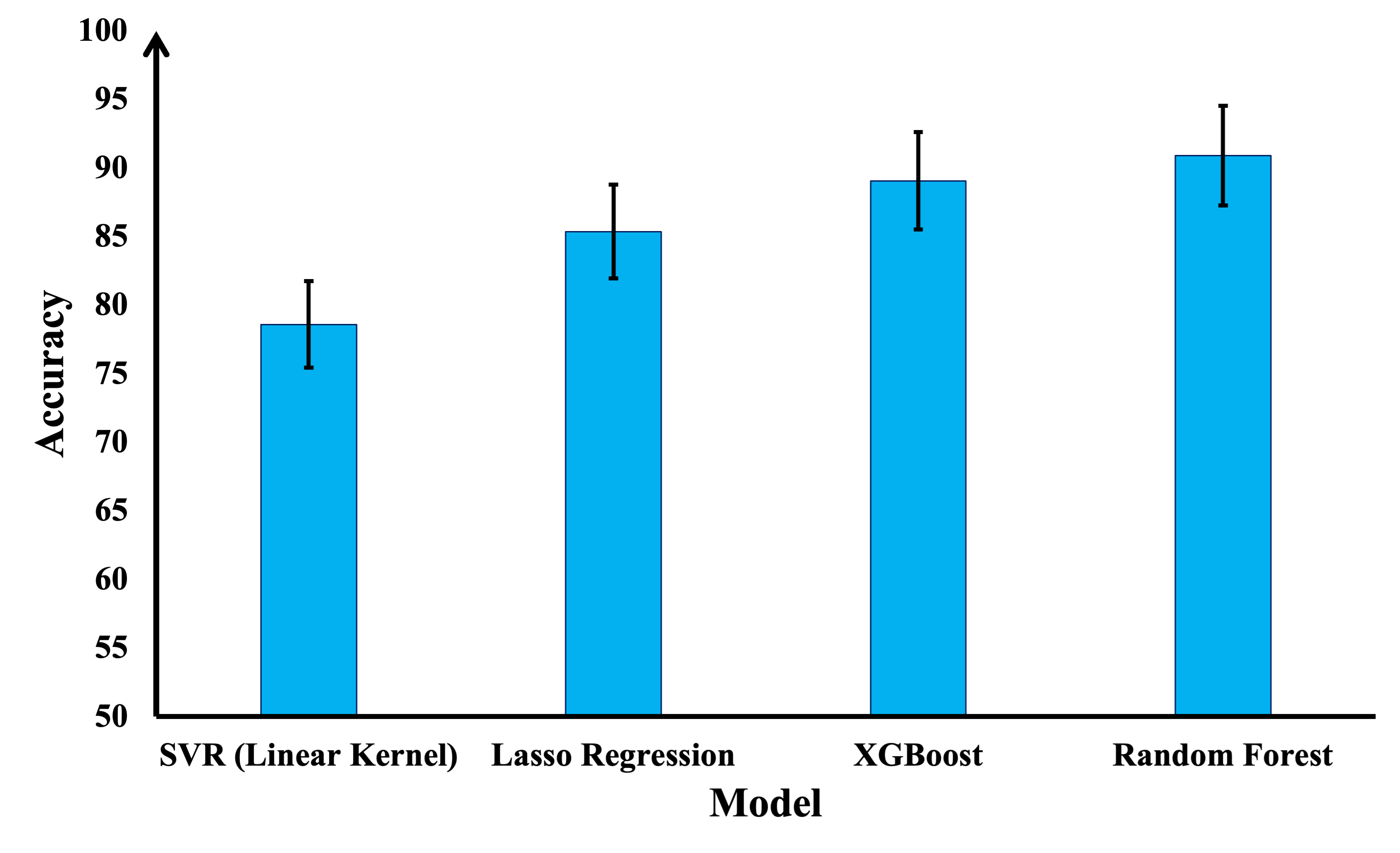}
    \caption{Comparison of model accuracy for predicting YS}
    \label{fig:accuracy}
\end{figure}

Implant materials must be able to withstand operational stresses, such as weight-bearing and movement, and adequate YS is one of the key properties that define this resistance. Four different models were constructed and implemented on the collected dataset to predict the YS of Mg alloys. Fig. \ref{fig:accuracy} compares the accuracy of these four models. After undergoing hyperparameter optimization, the RF model achieved a relatively high accuracy level of 91\% in predicting YS on 20\% of the unseen test sets. We used this model as the fitness function in a genetic algorithm. XGBoost followed RF in performance with 89\% accuracy on the unseen test set.

\subsection{Simulation Validation}
\begin{figure*}[htb]
    \centering
    \includegraphics[width=1\linewidth]{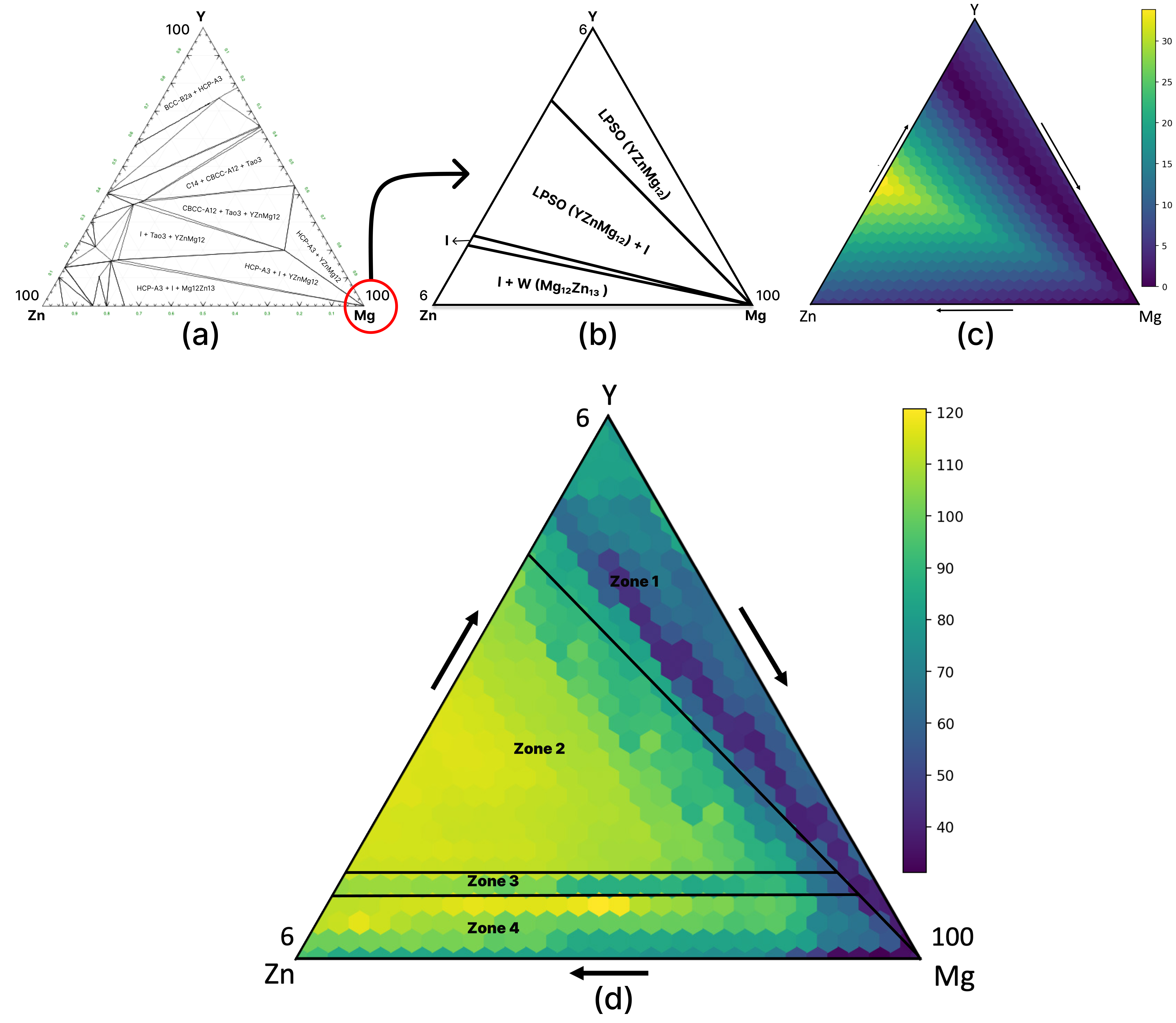}
    \caption{a) Ternary Mg-Zn-Y phase diagram calculated by FactSage; b) Mg-rich corner of the Mg-Zn-Y phase diagram; c) total amounts of second phases map of the Mg-rich corner d) YS map of the Mg-rich corner of Mg-Zn-Y alloy system.}
    \label{fig:ternary}
\end{figure*}

To validate the predictive model, we constructed a YS map of ternary alloys using the RF model. The map consists of a gridline with numerous points, each with three quantities: the amounts of the alloying elements (Zn and Y in the ternary map of fig. \ref{fig:ternary}) and Mg as the main element. The RF model predicted the YS of each point on the gridline, and the ternary map was created accordingly. As an example, fig. \ref{fig:ternary} (d) shows the ternary YS map of Mg-Zn-Y alloys with a 6\% limit for Zn and Y additions. The colour bar on the map represents the YS of the alloys in MPa, with yellow regions indicating higher YS and blue regions indicating lower YS.

It is evident from the YS ternary map (fig. \ref{fig:ternary} (d)) that there are four major YS zones. Regions 2 and 4 have high YS, while region 3 has moderate YS, and region 1 has low YS. The strength and quantity of different phases in the material influences the YS of metallic alloys. Phase diagrams display the equilibrium phases at various temperatures and for different elements in the alloys. Fig. \ref{fig:ternary} (a) shows the ternary phase diagram of Mg-Zn-Y alloys at room temperature, obtained using the CALPHAD technique from the software FactSage. The alloys considered are in the Mg-rich corner, with 6 wt\% being the upper limit for the level of alloying elements. This Mg-rich corner was used in the simulation, as shown in Figure \ref{fig:ternary} (b).

In cast alloys, YS depends on solid solution, grain boundary, and to a great extent, on second-phase strengthening. The type and amount of second phases are key factors. \citet{bai2020mechanical} confirmed that second-phase strengthening is the primary mechanism for increasing the mechanical properties of Mg-Zn-Y alloys.

Tab.~\ref{tab:ternary} summarizes the data from the YS ternary map (Fig.~\ref{fig:ternary} (d)), the phases from the ternary phase diagram for the four zones (Fig.~\ref{fig:ternary} (b)), and the total amounts of the second phases of the ternary phase diagram for the four zones (fig. \ref{fig:ternary} (c)). It is noteworthy that YS is well-correlated with the total amounts of the second phases for Zones 1-3, while the alloys of these zones have different phases of Long-period-stacking-order (LPSO) ($YZnMg_{12}$), or Quasicrystal I ($Mg_3Zn_6Y$), or both. Zone 4 has a different behaviour; the alloys in this zone have high YS despite having low to medium amounts of second phases. For Zone 4, the appearance of the W ($Mg_{12}Zn_{13}$) phase seems to have a considerable influence on YS. Quasicrystal I ($Mg_3Zn_6Y$) phase is also known to influence YS. \citet{ye2017microstructure} demonstrated that alloys containing the I phase in addition to the LPSO ($YZnMg_{12}$), or W ($Mg_{12}Zn_{13}$) phase would have a higher YS. \citet{xu2012effects} showed that an increase in the amount of I phase would increase the strength of Mg-Zn-Y alloys.

\begin{table}[htb]
\centering
\caption{Classification of the zones  in the Mg-Rich corner of Mg-Zn-Y system with respect to YS, and type and amount of second phases}
\label{tab:ternary}
\begin{adjustbox}{width=0.75\textwidth}
\begin{tabular}{|l|c|c|c|} 
\hline
Region & Range of YS & Second Phases & Total Amount of Second Phases \\ 
\hline
Zone 1 & Low & LPSO ($YZnMg_{12}$) & Low \\ 
\hline
Zone 2 & High & \begin{tabular}[c]{@{}c@{}}I ($Mg_3Zn_6Y$)\\LPSO ($YZnMg_{12}$)\end{tabular} & High\\ 
\hline
Zone 3 & Medium & I ($Mg_3Zn_6Y$) & Medium\\ 
\hline
Zone 4 & High & \begin{tabular}[c]{@{}c@{}}I ($Mg_3Zn_6Y$)\\W ($Mg_{12}Zn_{13}$)\end{tabular} & Low to medium\\
\hline
\end{tabular}
\end{adjustbox}
\end{table}

\subsection{Design of New Biocompatible Mg Alloys using the Genetic Algorithm (GA)}

\begin{figure}[h]
    \centering
    \includegraphics[width=0.48\linewidth]{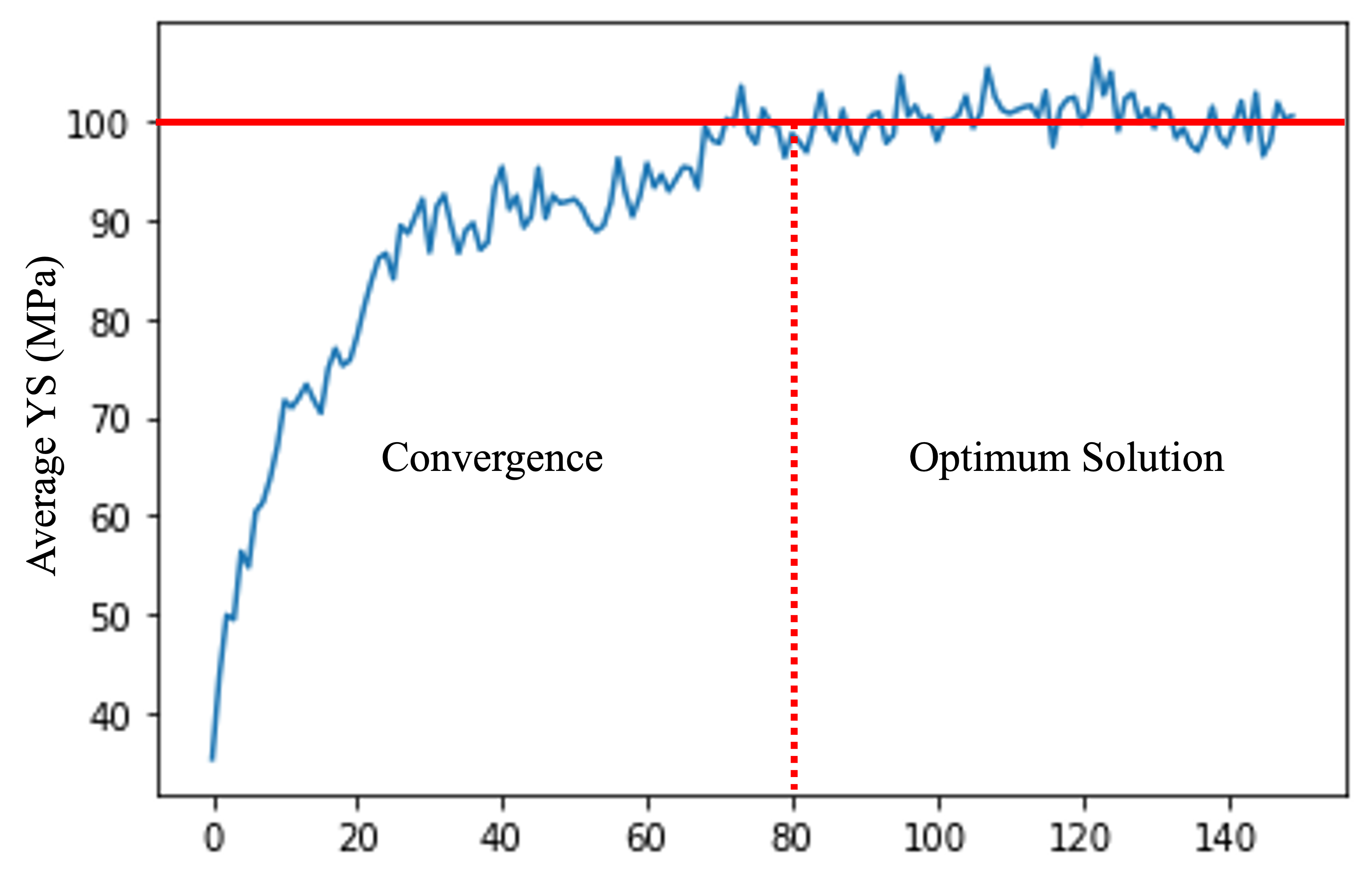} 
\caption{Average YS in each generation for alloys  that were not heat-treated, showing the optimization process wherein the average YS increased with each generation until optimized}
\label{fig:GA}
\end{figure}

This study aimed to identify an alloy composition that could match natural bone's hardness and YS while remaining biocompatible. Therefore, only biocompatible elements such as Zn, Ca, Sr, Mn, and Zr were considered in the genetic algorithm, and limits were set for adding each element for biocompatibility consideration. The limits for Zn, Ca, Sr, Mn, and Zr addition were 6, 2, 2, 0.75, and 1 wt\%, respectively. With each generation, the algorithm evolved and optimized the alloy composition. The genetic algorithm results were analyzed by monitoring each generation's average YS of the alloy compositions. As depicted in fig~\ref{fig:GA}, the average YS of the alloy compositions increased as the generations progressed. The initial randomly generated population had an average YS of approximately 35 MPa, which rose to approximately 100 MPa in the final population. This increase in YS indicates that the composition is nearing an optimal condition. However, to ensure that the algorithm did not become stuck on a global optimum, the genetic algorithm was run for several more iterations even after the composition appeared to have been optimized after 80 iterations.

Tab.~\ref{tab:Candidates} presents the compositions of two candidate alloys that were selected as potential biocompatible and biodegradable Mg alloys with improved strength. These candidates were chosen from the final iteration of the genetic algorithm based on their high YS. The RF model predicts that the YS of these candidates is 120 MPa for Alloy 1 and 129 MPa for Alloy 2, which is significantly higher than conventional biocompatible Mg alloys reported in the literature \citep{wang2016novel, bornapour2014magnesium}. The candidate alloys are predicted to have tensile YS slightly higher than that of natural bone but are expected to demonstrate compressive YS close to it (see Tab. \ref{tab:properties}).

FactSage calculations were performed to predict the phase selection in these alloys at equilibrium, and Scheil calculations were used to simulate the phases that would form under faster solidification conditions. The simulations predict a higher amount of second phases in cast Alloy 2 (4.92\%) than in cast Alloy 1 (3.29\%). Cast alloys usually have symmetric behaviour in tensile and compressive YS. Since the amount of second phases largely determines the YS, the trend in YS predicted for Alloy 1 and Alloy 2 by the GA has good scientific concordance. It is noted that based on Scheil calculation results, the main second phases in Alloy 2 are $Ca_{2}Mg_{55}Zn_{43}$ and $Sr_{2}Zn_{43}Mg_{55}$, while Alloy 1 mainly consists of $Sr_{18}Zn_{6}Mg_{20}$.

\begin{table}[htb]
\centering
\caption{Alloy Compositions (wt\%) suggested by GA and their predicted YS and the Phase selection predicted by FactSage}
\label{tab:Candidates}
\begin{adjustbox}{width=1\textwidth}
\begin{tabular}{|c|c|c|c|c|c|c|c|l|l|} 
\hline
\multirow{2}{*}{\begin{tabular}[c]{@{}c@{}}Candidate \\Alloys\end{tabular}} & \multirow{2}{*}{Mg} & \multirow{2}{*}{Ca~} & \multirow{2}{*}{Sr} & \multirow{2}{*}{Zn} & \multirow{2}{*}{Zr} & \multirow{2}{*}{Mn} & \multirow{2}{*}{\begin{tabular}[c]{@{}c@{}}Predicted YS \\(MPa)\end{tabular}} & \multicolumn{2}{c|}{\begin{tabular}[c]{@{}c@{}}Phases predicted by FactSage and\\~their amounts (in parentheses)\end{tabular}}                                                                                                                                                                                                                                                                                                                                          \\ 
\cline{9-10}
                                                                            &                     &                      &                     &                     &                     &                     &                                                                               & \multicolumn{1}{c|}{Schiel}                                                                                                                                                                                                                          & \multicolumn{1}{c|}{Equilibrium}                                                                                                                                                                                 \\ 
\hline
\multicolumn{1}{|l|}{Alloy 1}                                               & bal.                & 0.05                 & 0.5                 & 2.7                 & 0.75                & 0.3                 & 120                                                                           & \begin{tabular}[c]{@{}l@{}}\textbf{Major:}\\Mg-HCP-A3 (96.7), \\$Sr_{18}Zn_{6}Mg_{20}$ (1.73), \\$Sr_{2}Zn_{43}Mg_{55}$ (0.68). \\\textbf{Minor:}\\C14($Mn_{2}Zr$) (0.54), \\CaMgZn (0.25) \\\textbf{Total amount: 3.29}\end{tabular}                                               & \begin{tabular}[c]{@{}l@{}}\textbf{Major:}\\Mg-HCP-A3 (95.6),~\\$SrZn_{5}$ (2.26),\\C15 ($Zn_{2}Sr$) (1.22),~~\\\textbf{Minor:}\\C14 ($Mn_{2}Zr$) (0.54),~\\CaMgZn (0.25)\\\textbf{Total amount: 4.27}\end{tabular}                  \\ 
\hline
\multicolumn{1}{|l|}{Alloy 2}                                               & bal.                & 0.1                  & 0.1                 & 4.5                 & 0.02                & 0.15                & 129                                                                           & \begin{tabular}[c]{@{}l@{}}\textbf{Major:}\\Mg-HCP-A3 (95), \\$Ca_{2}Mg_{55}Zn_{43}$ (2.71) \\$Sr_{2}Zn_{43}Mg_{55}$ (1.73)~\\\textbf{Minor:}\\CaMgZn (0.18),~\\$Ca_{2}Mg_{5}Zn_{13}$ (0.11),~\\$Sr_{18}Zn_{6}Mg_{20}$ (0.11),\\Mn (0.04),\\C14 ($Mn_{2}Zr$) (0.04)~\\\textbf{Total amount: 4.92}\end{tabular} & \begin{tabular}[c]{@{}l@{}}\textbf{Major:~}\\Mg-HCP-A3 (93.5), \\$Ca_{2}Mg_{55}Zn_{43}$ (5.27)~~\\\textbf{Minor:}\\$Sr_{2}Zn_{43}Mg_{55}$ (0.68)\\$SrZn_{5}$ (0.34),~\\Mn (0.12), \\C14($Mn_{2}Zr$) (0.04) ~~\\\textbf{Total amount: 6.45}\end{tabular}  \\
\hline
\end{tabular}
\end{adjustbox}
\end{table}


\subsection{Experimental Validation}

Tab. \ref{tab:chemical} presents the chemical composition of the synthesized alloys. As observed from the table, there are slight variations in the composition of the suggested candidate alloys and the synthesized alloys, mainly in Zr and Mn. These elements have low recovery rates upon alloying, which could explain the differences observed.

\begin{table}[h]
\centering
\caption{Chemical composition and phase prediction of the synthesized alloys (wt\%)}
\label{tab:chemical}
\begin{adjustbox}{width=0.75\textwidth}
\begin{tabular}{|c|c|c|c|c|c|c|l|} 
\hline
Candidate & Mg   & Ca   & Sr   & Zn   & Zr   & Mn   & \multicolumn{1}{c|}{\begin{tabular}[c]{@{}c@{}}Phases predicted by FactSage (Scheil) \\and their~amounts~(in parentheses)\end{tabular}}                                                                      \\ 
\hline
Alloy 1   & bal. & 0.05 & 0.5 & 2.8  & 0.05 & 0.2 & \begin{tabular}[c]{@{}l@{}}\textbf{Major:}~\\Mg-HCP-A3 (96.3), \\$Sr_{2}Zn_{43}Mg_{55}$ (1.9),~\\$Sr_{18}Zn_{6}Mg_{20}$ (1.3)\\\textbf{Minor:}~\\CaMgZn(0.31), \\C14 ($Mn_{2}Zr$) (0.1)\\\textbf{Total amount: 3.63}\end{tabular}         \\ 
\hline
Alloy 2   & bal. & 0.1 & 0.1  & 5.4 & 0.01    & 0.15 & \begin{tabular}[c]{@{}l@{}}\textbf{Major:}~\\Mg-HCP-A3 (94), \\$Ca_{2}Zn_{43}Mg_{55}$~(3.6),\\$Sr_{2}Zn_{43}Mg_{55}$(2.2)\\\textbf{Minor:}~\\CaMgZn(0.07),~\\$Ca_{2}Mg_{5}Zn_{13}$ (0.04),~\\Mn(0.01)\\\textbf{Total amount: 5.95\%}\end{tabular}  \\
\hline
\end{tabular}
\end{adjustbox}
\end{table}

\subsubsection{Second phases analysis}
The phases in these alloys were predicted using FactSage and are presented in Tab. \ref{tab:chemical}. Scheil calculations assume no-back diffusion in the solid and are used to assess phase selection under medium solidification rates, such as permanent-mould casting used in producing the alloy samples. A comparison of the FactSage Scheil calculations for the candidate alloys (presented in tab. \ref{tab:Candidates}) and the synthesized alloys (presented in tab. \ref{tab:chemical}) shows that the major phases in the alloys are pretty similar, with slight changes in their amounts.

\subsubsection{Microstructure analysis}

\begin{figure*}[h]
\centering
    \begin{subfigure}{0.4\textwidth}
        \centering
        \includegraphics[width=0.8\linewidth]{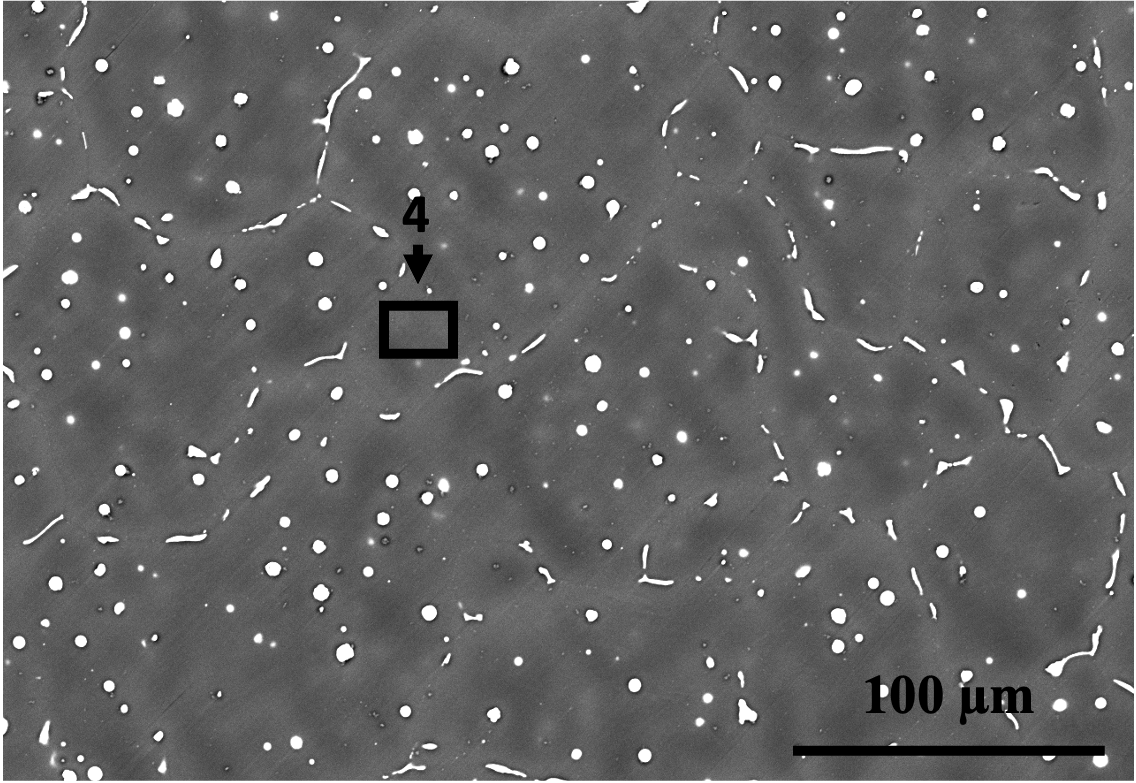}  
        \caption{}
        \label{fig:A2}
    \end{subfigure}
    \begin{subfigure}{0.4\textwidth}
        \centering
        \includegraphics[width=0.8\linewidth]{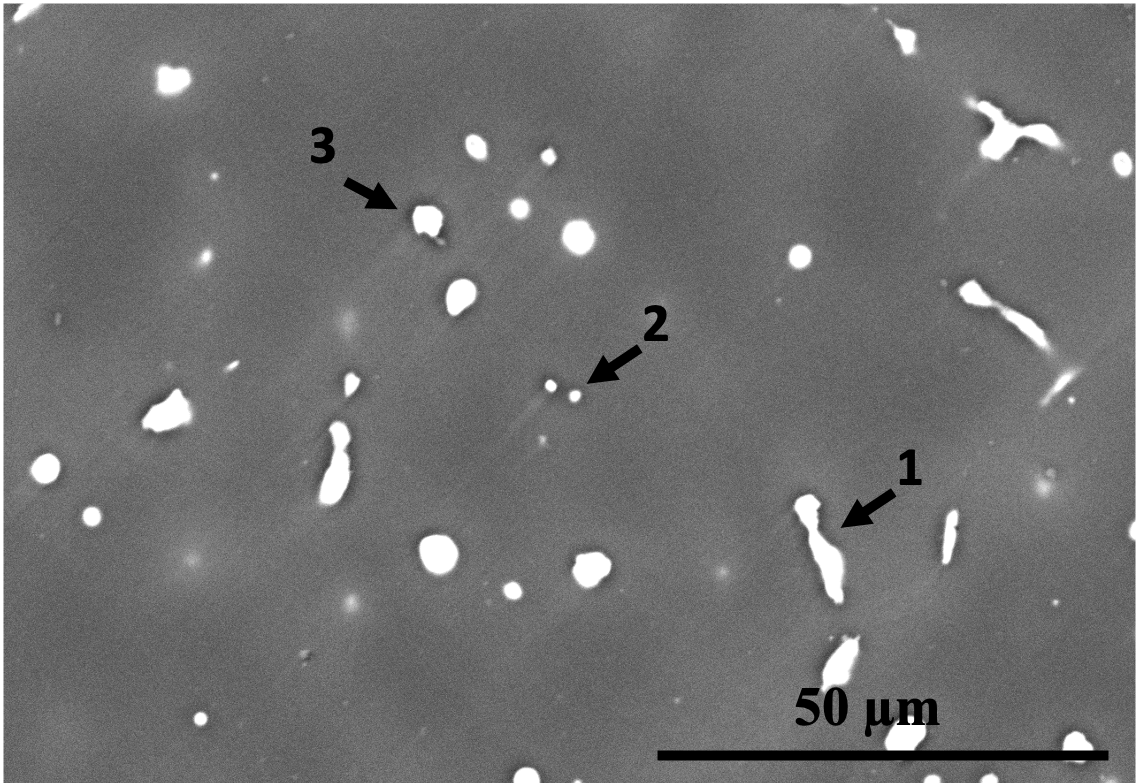}  
        \caption{}
        \label{fig:A2-zoom}
    \end{subfigure}
    \begin{subfigure}{0.4\textwidth}
        \centering
        \includegraphics[width=0.8\linewidth]{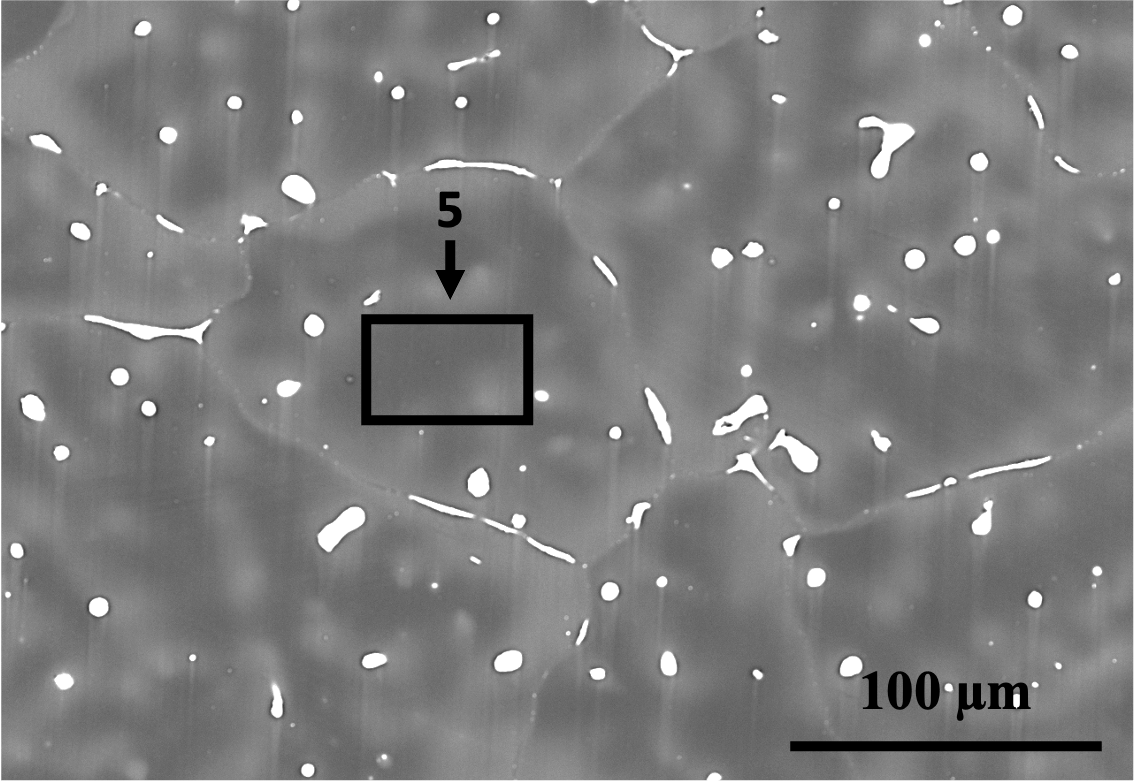}  
        \caption{}
        \label{fig:A1}
    \end{subfigure}
    \begin{subfigure}{0.4\textwidth}
        \centering
        \includegraphics[width=0.8\linewidth]{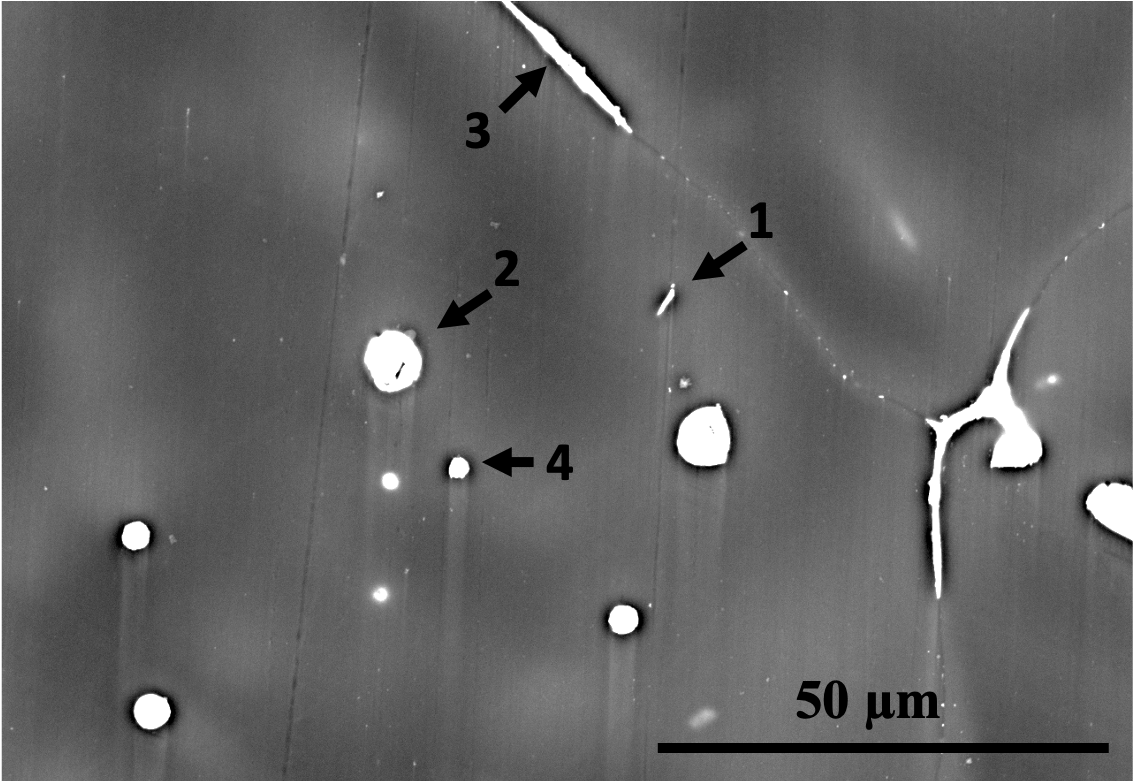}  
        \caption{}
        \label{fig:A1-zoom}
    \end{subfigure}
\caption{SEM-BSE images from the longitudinal cross sections a \& b) Alloy 1 and c \& d) Alloy 2. Regions marked with numbers are spots of EDS analysis (see Tab. \ref{tab:EDS})}
\label{fig:SEM}
\end{figure*}

The microstructures of the two alloys are displayed in Fig. \ref{fig:SEM}. It is evident from the figures that Alloy 1 has a finer microstructure (i.e., smaller dendrite arm and grain size) compared to Alloy 2. Despite the fact that Alloy 2 has a higher Zn content, which is known to decrease grain size in Mg alloys \citep{becerra2009effects}, Alloy 1 has finer grains due to the presence of Zr and Mn. Zr and Mn are known to have a grain-refining effect, with Zr being the most effective grain refiner for Mg-based alloys lacking Al, such as Mg-Zn and Mg-RE \citep{chen2014recent}. However, in Al-containing alloys, Zr reacts with Al.

The SEM images in Fig. \ref{fig:SEM} and the EDS analyses in Tab. \ref{tab:EDS} reveal the presence of second phases in both alloys composed of Ca, Sr, Zn, and/or Mn. The phases in the microstructures of the two alloys were identified using SEM/EDS analysis combined with the predictions of FactSage simulations, as shown in Tab. \ref{tab:EDS}. The limitations of SEM/EDS due to signal from the interaction volume were also considered. It can be observed that the second phases in Alloy 1 are finer in comparison to those in Alloy 2. This is partly because the grains and dendrites are refined, resulting in more dispersed intergranular/interdendritic phases due to increased boundary area. Additionally, the existence of Zr and higher amounts of Mn in Alloy 1 contributes to the finer second phase particles. Mn is a peritectic element that is enriched intradendritically and can have a co-precipitating/nucleating role for other precipitates \citep{celikin2012effect}, as seen in spot 2 in Fig. \ref{fig:A2-zoom}.

\begin{table}[hb]
\caption{EDS analysis of the spots shown in Fig. \ref{fig:SEM}}
\label{tab:EDS}
\centering
\begin{adjustbox}{width=0.86\textwidth}
\begin{tabular}{|c|c|c|c|c|c|l|} 
\hline
Zone & Mg & Ca  & Sr  & Zn & Mn & \multicolumn{1}{c|}{Phase identification based on EDS and FactSage}  \\ 
\hline
\multicolumn{7}{|c|}{\textbf{\textit{Alloy 1}}}                                                                  \\ 
\hline
1    & 96 & -   & 0.5 & 4  & -  & $Sr_{2}Zn_{43}Mg_{55}$                                                                              \\ 
\hline
2    & 93 & -   & -   & 5  & 2  & Mn and a Zn-bearing phase                                                                \\ 
\hline
3    & 96 & 0.1 & -   & 4  & -  & $Ca_{2}Mg_{55}Zn_{43}$                                                                               \\ 
\hline
4    & 98 & -   & -   & 2  & -  & \multicolumn{1}{l|}{Mg(Zn) solid solution}                                                \\
\hline
\multicolumn{7}{|c|}{\textbf{\textit{Alloy 2}}}                                                                 \\ 
\hline
1    & 87 & -   & -   & 5  & 8  & C14 ($Mn_{2}Zr$) and signal from the $alpha-Mg(Zn)$                                                \\ 
\hline
2    & 40 & 5   & 8   & 48 & -  & $Sr_{2}Zn_{43}Mg_{55}$ or $Sr_{18}Zn_{6}Mg_{20}$ and CaMgZn                                                     \\ 
\hline
3    & 72 & 1   & -   & 27 & -  & CaMgZn                                                                                    \\ 
\hline
4    & 93 & -   & -   & 7  & -  & Mg(Zn) solid solution and a Zn-bearing phase                                             \\ 
\hline
5    & 97 & -   & -   & 3  & -  & Mg(Zn) solid solution                                                                     \\ 
\hline
\end{tabular}
\end{adjustbox}
\end{table}

\subsubsection{Mechanical property analysis}

The hardness measurements indicate Alloy 2 has a Vickers hardness of 52HV (52 kgf/$mm^2$=49 BHN) and Alloy 1 has a Vickers hardness of 47 HV (47 kgf/$mm^2$=45BHN). Comparing the YSs and BHN of Mg cast alloys shows that the YS in megapascals (MPa) is, on average, 2.4 BHN. Therefore, the YSs of the synthesized alloys can be estimated as 113 MPa for Alloy 2 and 108 MPa for Alloy 1.

\begin{figure}[h]
    \centering
    \includegraphics[width=0.7\linewidth]{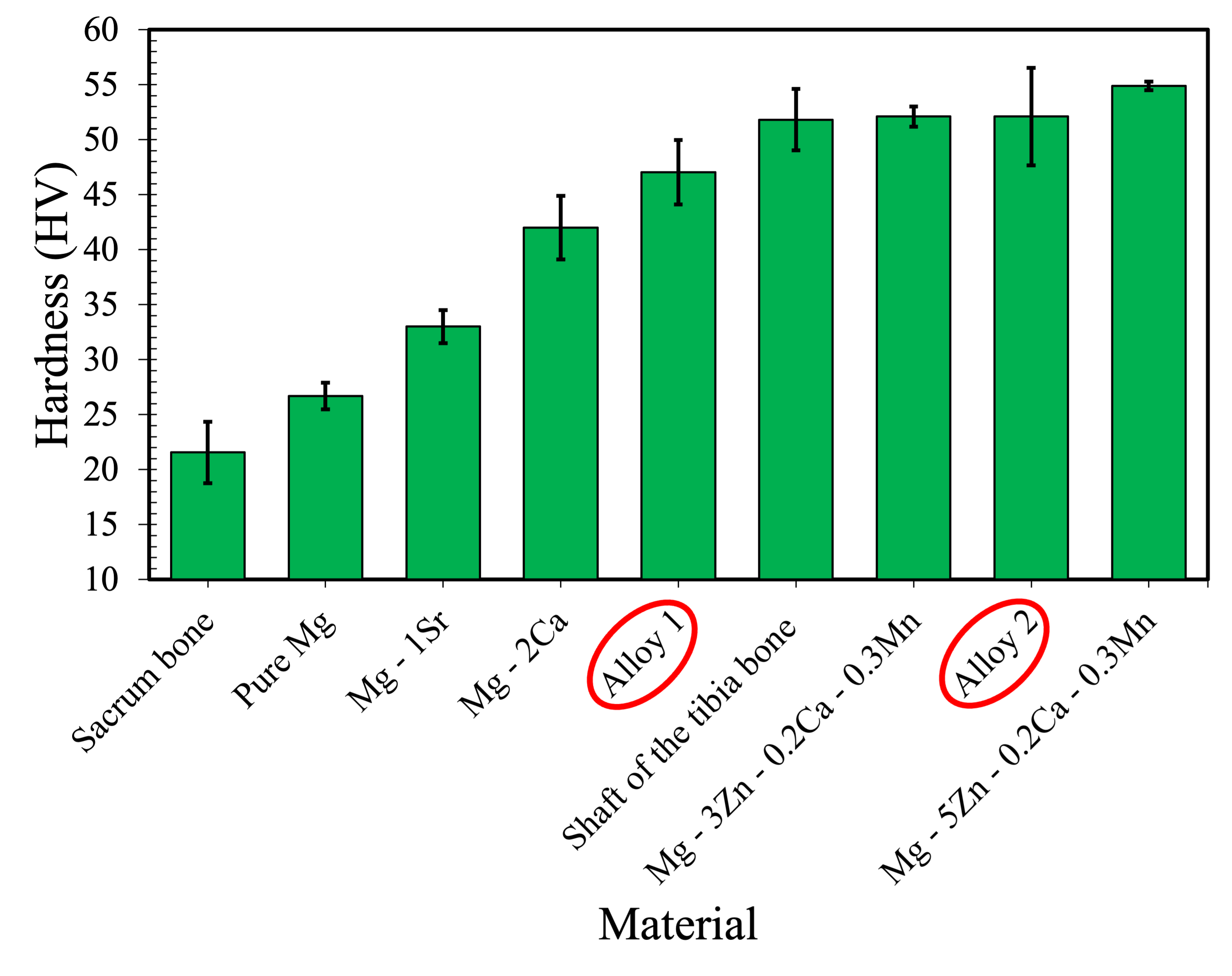} 
\caption{Comparison of hardness of human bones \citep{li2021atlas}, Mg-Ca \citep{harandi2013effect}, Mg-Sr \citep{brar2012investigation}, Mg-Ca-Zn alloys \citep{gungor2021effects}, and the synthesised alloys}
\label{fig:hardness}
\end{figure}

The synthesized alloys' hardness is compared with that of other biocompatible Mg alloys and natural bone in Fig.~\ref{fig:hardness}. The materials depicted include the synthesized alloys of this study, natural bone, and other experimental biodegradable Mg alloys. The results indicate that synthesized Alloy 2, with a value of 52 HV (52$kgf/mm^2$=49 BHN), exhibits a higher level of hardness among the two synthesized alloys and is similar to other experimental alloys and to the shaft of the tibia bone. It is worth noting that the synthesized alloys also have a hardness level that is close to that of natural bone, which will prevent stress shielding.

The experimental alloys and the suggested compositions by the model have similar YSs. However, since the compositions of the experimental alloys were slightly different from the suggested compositions, the model was used to predict the actual compositions' YSs. The predicted YS of experimental Alloy 1 was 118 MPa, and that of experimental Alloy 2 was 125 MPa, which follows a similar trend as the experimental results. Furthermore, microstructural analysis of the experimental alloys reveals that Alloy 1 has a finer dendrite arm spacing (Fig. \ref{fig:SEM}), while Alloy 2 has slightly higher hardness and YS than Alloy 1.

It should be noted that the hardness measurements of both experimental alloys show substantial deviation, with Alloy 2 displaying a more significant variation. The correlation between YS and hardness in metals often carries a 10-15\% error.

\section{Conclusion and future work}
This study has provided valuable insights into the potential of using advanced computational techniques, such as a combination of an RF model and a genetic algorithm, to optimize the composition of high-strength, biocompatible alloys for bio-implant applications. By taking into account only alloying elements that have been previously established as biocompatible and setting specific limits for each element, the algorithm was able to identify two candidate alloys with a high predicted YS that closely resemble the mechanical properties of natural bone. These alloys present themselves as promising solution for bio-implant applications that require both high strength and biocompatibility. However, it is essential to note that further research is necessary to fully validate these findings through mechanical testing of the alloys, as well as in-vivo testing to assess their biocompatibility and performance within a living organism. It is also essential to continue to optimize the model to embrace many aspects of a magnesium-based implant alloy with adequate biocompatibility, biodegradability, and mechanical behaviour. Continuing the materials science-based feature engineering is needed develop an integrated model for the design of Mg bio-implant alloys that can include the solid-solubility of the Mg matrix for predicting YS, image recognition to add microstructural features that govern UTS and the galvanic potential of the phases for assessing bio-corrosion resistance. Overall, this study highlights the potential for using advanced computational techniques to aid in developing new materials for bio-implant applications and opens the door for future research in this field.

\section{Acknowledgements}
This project was conducted under the Natural Sciences and Engineering Research Council (NSERC) of Canada Discovery Grant. One of the authors, Parham Valipoorsalimi, acknowledges the financial support of the McGill Excellence Fellowship during his Master's.

\section{Data Availability}
The dataset and code used in this study will be available upon request.
\section{Conflict of Interest}
The authors declare that they have no conflict of interest.

\bibliographystyle{elsarticle-num-names} 
\bibliography{cas-refs}





\end{document}